# Early Warning Signs of the Economic Crisis in Greece: A Warning for Other Countries and Regions

Ron W Nielsen[1]

Environmental Futures Research Institute, Gold Coast Campus, Griffith University, Qld, 4222, Australia

November, 2015

Warning signs about the developing economic crisis in Greece were present in the growth rate of the Gross Domestic Product (GDP) and in the growth of the GDP well before the economic collapse. The growth rate was strongly unstable. On average, in less than 50 years, it decreased 10-folds but after reaching a low minimum it quickly increased 6-folds only to crash before completing the full cycle. The decreasing growth rate was leading to an asymptotic maximum of the GDP but it was soon replaced by a fast-increasing growth rate propelling the GDP along a pseudo-hyperbolic trajectory, which if continued would have escaped to infinity in 2017. Such a growth could not have been possibly supported. Under these conditions, the economic collapse in Greece was inevitable.

**Introduction**

Factors influencing economic growth in Greece, both endogenous and exogenous have been extensively discussed (Felton & Reinhart, 2008; Katsimi & Moutos, 2010; Kotios & Pavlidis, 2012; Lapavitsas, 2009; Lyrintzis 2011; Oltheten, Pinteris & Sougiannis, 2003; Petrakos, Fotopoulos & Kallioras, 2012; US Senate 2011). Here we are going to show how a study of the growth rate of the Gross Domestic Product (GDP) can help to recognise early warning signs about potential economic crisis. Such warning signs are now ubiquitous, including the warning signs about the future of the global economic growth (Nielsen, 2015a). Our discussion is supported by the World Bank data (World Bank, 2015).

**Mathematical methods**

Mathematical methods of analysis of any type of growth have been described earlier (Nielsen, 2015b). Such investigations usually focus on the direct analysis of growing entities. However, it has been pointed out that the analysis of data can be extended by including also the analysis of the empirically-determined growth rates, $R$, which can be represented either as a function of time

$$R = \frac{1}{S}\frac{dS}{dt} = f(t) \tag{1}$$

or as a function of the size $S$ of the growing entity

---

[1] AKA Jan Nurzynski, r.nielsen@griffith.edu.au; ronwnielsen@gmail.com; http://home.iprimus.com.au/nielsens/ronnielsen.html

$$R = \frac{1}{S}\frac{dS}{dt} = f(S). \qquad (2)$$

In the first case, the solution of the differential equation (1) can be expressed using the general formula.

$$S(t) = \exp\left[\int f(t)dt\right]. \qquad (3)$$

In the second case, there is no such general description and different methods have to be used to solve the eqn (2). Furthermore, the eqn (2) has to be often solved numerically.

Empirical growth rate can be determined using the gradient calculated directly from data or by polynomial interpolation. It is generally essential to use interpolated gradient because in order to reproduce the past growth and to predict the future we have to understand the growth-rate trajectory.

**Analysis of the economic growth in Greece**

Growth rates of the GDP in Greece between 1960 and 2014 calculated using the World Bank data (World Bank, 2015) are shown in Figure 1. As expected, the growth rate calculated directly from data, $R(Direct)$, shows strong fluctuations, which are the reflection of the fluctuating gradient. They disappear when the interpolated gradient is used and a clear trend is then revealed.

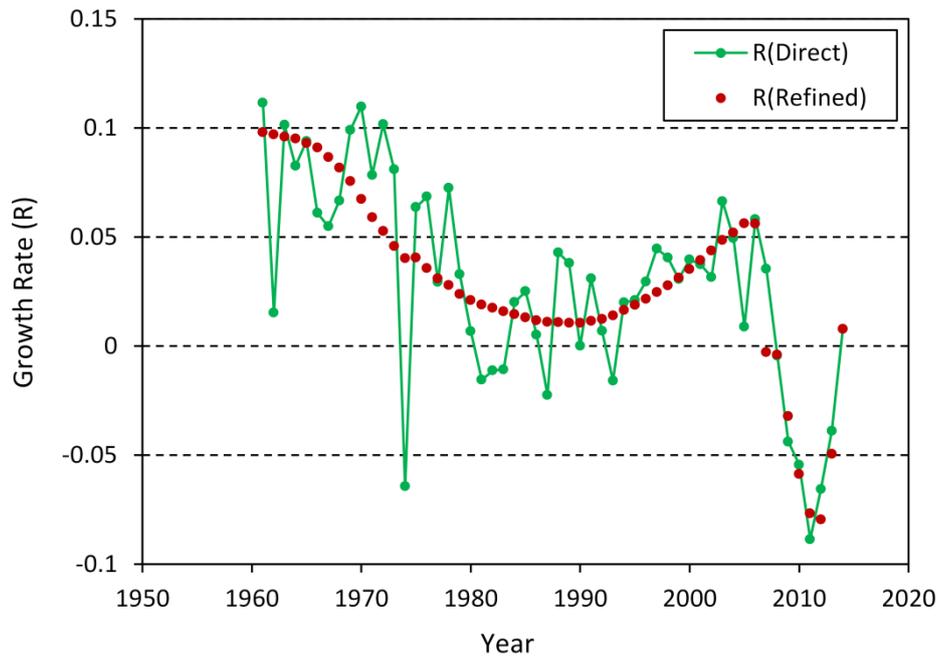

Figure 1. Empirical growth rate of the Greek GDP calculated directly from the GDP data (World Bank, 2015) [$R$ ($Direct$)] and by using interpolated gradient [$R$ ($Refined$)].

The general trend of the growth rate of the GDP is characterised by changes over a large range of values. On average, the growth rate decreased about 10-folds in a relatively short time and increased about 6-folds. However, it crashed before completing its full cycle.

Small oscillations in the growth rate can be tolerated but strong variations are undesirable. Variations of this kind characterise, for instance, the economic growth in Sub-Saharan Africa, as shown in Figure 2. The general trend, which can be seen also in the growth rate calculated directly from data, was a fast decrease followed by a fast increase. However, unlike the growth rate characterising the economic growth in Greece, the strongly-varying growth rate in Sub-Saharan Africa completed it full cycle without yet causing economic collapse. The amplitude of the growth-rate oscillations was also smaller.

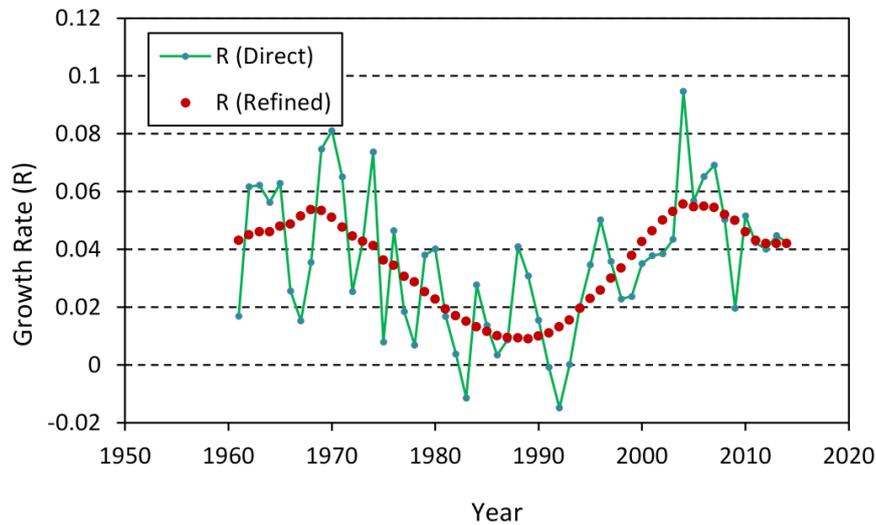

Figure 2. The strongly-changing growth rate of the GDP in Sub-Saharan Africa resembles the strongly-changing growth rate in Greece before the economic collapse.

Signs of developing crisis in Greece can be seen even more clearly if we examine the dependence of the empirical growth rate on the *size* of the GDP, as shown in Figure 3. For better clarity we present only the size-dependence of the refined growth rate. Here the pattern is easier to analyse because it is represented by straight lines.

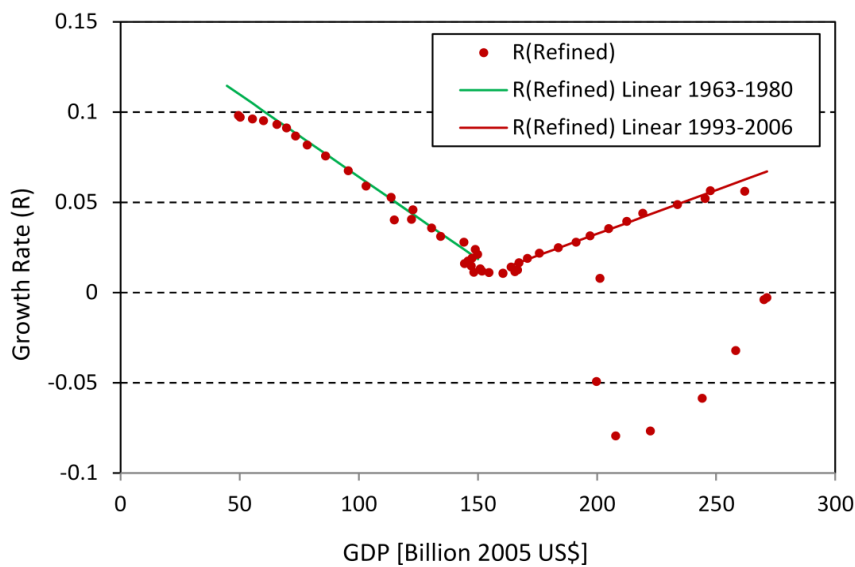

Figure 3. Growth rate of the GDP in Greece, plotted as a function of the *size* of the GDP, shows two clear linear trends. The points below the increasing linear trend are a signature of the strong reversal of the economic growth.

At first, the growth rate was decreasing linearly with the size of the GDP. Such a trend is desirable because it leads to an asymptotic maximum. However, the growth rate was decreasing too rapidly. It soon reached an intolerably low value and became unstable. At this stage, in order to maintain the previous logistic growth the growth rate should have continued to decrease linearly with the size of the GDP but very slowly. To control such a growth is difficult, perhaps even impossible, because the exceedingly slowly decreasing growth rate can be easily constant generating exponential growth or it can get out of control and start to increase, propelling the economic growth along a fast-increasing trajectory. This is precisely what happened in Greece. After a brief period of instability, the growth rate started to increase.

Any economic growth characterised by a consistently increasing growth rate should be, if possible, avoided. In particular, economic growth characterised by the growth rate increasing linearly with the size of the GDP follows a pseudo-hyperbolic trajectory (Nielsen, 2015b) containing singularity and thus escaping to infinity at a fixed time. Every effort should be made to discontinue such a growth as quickly as possible.

Growth rate decreasing or increasing linearly with the size of the growing entity, $S(t)$, in our case the GDP, generates growth trajectory described by the following equation (Nielsen, 2015b):

$$S(t) = \left[ Ce^{-at} - \frac{b}{a} \right]^{-1}. \qquad (5)$$

For the growth rate *decreasing* with the size $S(t)$ (as shown in Figure 3 for the GDP between around \$50 billion and \$150 billion) the parameter $b < 0$ and the eqn (5) describes the logistic-type of growth, which approaches asymptotically the limit of $a/|b|$. Logistic limit is often called incorrectly the carrying capacity. It is just the calculated limit for a certain set of parameters $a$ and $b$, which are not necessarily describing the *ecologically*-defined limit to growth.

For the growth rate *increasing* with the size $S(t)$, the parameter $b > 0$ and the eqn (5) describes a pseudo-hyperbolic growth escaping to infinity at the time

$$t_s = -\frac{1}{a} \ln \frac{b}{aC}. \qquad (6)$$

The distinction between the pseudo-hyperbolic and hyperbolic distributions is in the non-essential details (Nielsen, 2015b). Effects for their positive values are the same. They have similar shape and they both escape to infinity at a fixed time. However, while the reciprocal values of the hyperbolic distribution follow a decreasing straight line, the reciprocal values of the pseudo-hyperbolic distribution described by the eqn (5) for $b > 0$ decrease non-linearly.

Parameters describing the linearly-*decreasing* growth rate of the GDP are $a = 1.553 \times 10^{-1}$ and $b = -9.112 \times 10^{-4}$. For this empirically-determined set of parameters, the asymptotic limit to growth is around \$170 billion (of 2005 US\$).

Parameters describing the linearly-*increasing* growth-rate of the GDP in Greece are: $a = -6.424 \times 10^{-2}$ and $b = 4.839 \times 10^{-4}$. For these empirically-determined parameters the economic growth in Greece was now following a hazardous, pseudo-hyperbolic trajectory with a singularity, which was not located in a distant future but just on the doorstep, in 2017. The economic growth in Greece came very close to this impossible-to-reach point and inevitably collapsed.

The GDP trajectories associated with the linearly-decreasing and linearly-increasing growth rate are shown in Figure 4.

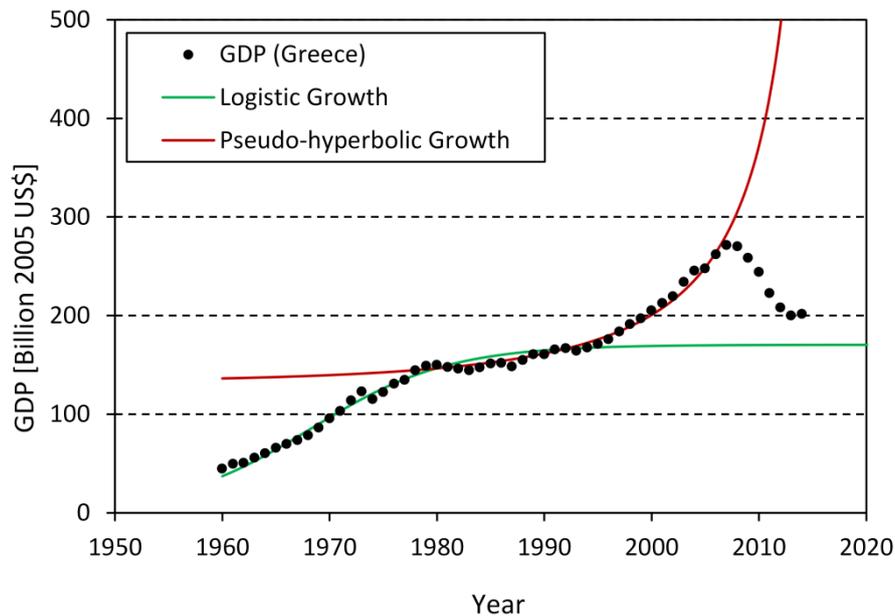

Figure 4. Economic growth in Greece and the GDP trajectories calculated using the eqn (5) and parameters *a* and *b* determined by fitting linear trends to the GDP-dependent growth rate shown in Figure 3.

The GDP in Greece increased from $44.7 billion (2005 US$) in 1960 to $271 billion in 2007, representing a 6-fold increase in 47 years. To maintain the momentum of the economic growth corresponding to the rapidly-increasing growth rate, which propelled the economic growth along the fast-increasing pseudo-hyperbolic trajectory, the same increase in the GDP would have to be generated in just 5 years, and then, in just additional 5 years the GDP would have to be pushed to infinity. Such a growth was simply impossible and it was bound to collapse.

Warning signs about the developing economic crisis in Greece were clearly displayed as early as between 1960 and 1970. The growth rate was *decreasing too fast*. After reaching its *low minimum* there was a certain degree of *instability*, which can be seen clearly by studying the dependence of the growth rate on the size of the GDP. The growth rate then started to increase but this time it was *increasing too fast*. Furthermore, even in the early stages of this fast increase it was *increasing linearly* with the size of the GDP. Consequently, the GDP was increasing along a fast-increasing *pseudo-hyperbolic trajectory* characterised by *singularity*. If properly analysed, this linear trajectory of the growth rate could have indicated *a close proximity of singularity*. It would then have been obvious that such a rapid increase of the growth rate could not be tolerated.

Now, after completing a loop (see Figure 3) the growth rate and the GDP were pushed about 25 and 15 years back in time, respectively. From now on, it appears that the growth rate has nowhere to go. There is little room for a safe manoeuvre.

The growth rate cannot be zero because a zero growth rate would generate a constant GDP but the GDP is intolerably low. If the growth rate is going to remain constant, the growth will be exponential. Over a long time, exponential growth is undesirable but for a small growth rate it might be tolerated. However, the generated annual increase in the GDP might be hard to accept. So, it is quite possible, that the growth rate will start to increase again, perhaps even along a linear GDP-dependent trajectory, which would push Greek economy towards a new economic crisis.

A safe option (if the economic growth in Greece is not going to be guided by chance but by options) would be to increase the growth rate to a certain safe level and then start to decrease it linearly but gently with time. Such an option would gradually boost the growth of the GDP and would maintain its growth along a safe trajectory.

**Conclusions**

Close analysis of growth rates of the Gross Domestic Product can be used to identify early signs of economic crisis. Growth-rate data can be used to predict growth of the GDP and to investigate its security. Such information can then be used to plan early intervention strategies to avoid a potential economic collapse.

We have analysed economic growth in Greece and we have shown that there were ample warning signs about the economic insecurity. Such signs, when early recognised, could have perhaps helped to stabilise the Greek economy.

The future of the economic growth in Greece is still uncertain. A safe option appears to be still possible but it is also possible that the future economic growth will be controlled by chance. A general guiding principle for any economic growth at the current stage of economic activities, global, regional, or national, is to avoid a constant growth rate or the growth rate increasing with time or with the size of the GDP, because such growth rates are associated with unsustainable economic growth. The preferable growth rate trajectory is a slow linear decrease.